\documentclass[12pt]{iopart}
\usepackage{iopams}
\usepackage{amssymb}
\usepackage{graphicx,subfigure,epstopdf}

\usepackage{cite}
\eqnobysec
\usepackage[colorlinks,linktocpage,linkcolor=blue]{hyperref}

\begin{document}

\title[]{Tempered fractional Langevin-Brownian motion with inverse $\beta$-stable subordinator}

\author{Yao Chen, Xudong Wang, and Weihua Deng}

\address{School of Mathematics and Statistics, Gansu Key Laboratory of Applied Mathematics and Complex Systems, Lanzhou University, Lanzhou 730000, P.R. China}
\ead{ychen2015@lzu.edu.cn, xdwang14@lzu.edu.cn, and dengwh@lzu.edu.cn}
\vspace{10pt}

\begin{abstract}
Time-changed stochastic processes have attracted great attention and wide interests due to their extensive applications, especially in financial time series, biology and physics.
This paper pays attention to a special stochastic process, tempered fractional Langevin motion, which is non-Markovian and undergoes ballistic diffusion for long times.
The corresponding time-changed  Langevin system with inverse $\beta$-stable subordinator is discussed in detail, including its diffusion type, moments, Klein-Kramers equation, and the correlation structure. Interestingly, this subordination could result in both subdiffusion and  superdiffusion, depending on the value of $\beta$. The difference between the subordinated tempered fractional Langevin equation and the subordinated Langevin equation with external biasing force is studied for a deeper understanding of subordinator. The time-changed tempered fractional Brownian motion by inverse $\beta$-stable subordinator is also considered, as well as the correlation structure of its increments. Some properties of the statistical quantities of the time-changed process are discussed, displaying striking differences compared with the original process.




\noindent (Some figures may appear in colour only in the online journal)
\end{abstract}

\section{Introduction}
Over the last two decades, great efforts have been devoted to the study of anomalous diffusion phenomenon \cite{Hughes:1995,Metzler:2000}, which is ubiquitous in the natural world.
The specific form of anomalous diffusion is determined by the complex environment, modelled by the stochastic processes.
It is in general characterized by the mean squared displacement (MSD), i.e., $\langle (\Delta x(t))^2\rangle=\langle[x(t)-\langle x(t)\rangle]^2\rangle$. For the normal diffusion, MSD exhibits linear time dependence, $\langle (\Delta x(t))^2\rangle\simeq t$. For the anomalous diffusion, MSD has the non-linear time dependence $\langle (\Delta x(t))^2\rangle\simeq t^\alpha$ with $\alpha\geq 0$ and $\alpha\neq 1$. More specifically, we call it subdiffusion if $0<\alpha<1$ and superdiffusion if $\alpha>1$. For the special cases $\alpha=0$, $\alpha=2$, and $\alpha=3$,
the anomalous diffusion phenomena are, respectively, called localization, ballistic diffusion, and turbulent-diffusion.

For some data in real life, such as biology \cite{Golding:2006}, financial time series \cite{Janczura:2011}, ecology \cite{Scher:2002}, and physics \cite{Nezhadhaghighi:2011}, the time-changed stochastic process is needed, where the deterministic time variable is replaced by a positive non-decreasing random process and thus a combination of two independent random processes is produced.
One of the processes is called external process (or the original process), and another one is called internal process (or a subordinator). The idea of subordination was put forward by Bochner \cite{Bochner:1949} in 1949.
In recent years, the time-changed stochastic processes by subordinator or inverse subordinator have been widely discussed. For example, the path properties of the subordinated Brownian motion (Bm) have been investigated in \cite{Magdziarz:2010}; the covariance function and the Fokker-Planck equation of the time-changed Ornstein-Uhlenbeck process have been shown in \cite{Gajda:2015}; in \cite{Fogedby:1994,Meerschaert:2004}, it was displayed that the time-changed L\'{e}vy process by inverse stable subordinator is a limit process of the continuous time random walk (CTRW) models with the power law distributed random waiting times between each random jump, and \cite{Leonenko:2014} showed the correlation structure of time-changed L\'{e}vy process; besides that, Ref. \cite{Mijena,Kumar:2017,Wylomanska:2016,Hahn:2011} considered the time-changed fractional Brownian motion (fBm), discussing the moments, correlation structure, Fokker-Planck equation, and so on.

For convenience of discussion, we briefly review the definitions and properties of the subordinator as well as its inverse process. Subordinator, denoted as $t(s)$ here, which could be thought as a stochastic model of time evolution, is a non-decreasing L\'{e}vy process with stationary and independent increments  \cite{Applebaum:2009}. The first-passage time of a subordinator $\{t(s),s\geq 0\}$ is called inverse subordinator $\{s(t),t\geq 0\}$ \cite{Kumar:2015,Alrawashdeh:2017}, defined as
\begin{equation}
s(t)=\inf_{s>0}\{s:t(s)>t\}.
\end{equation}
Let $t(s)$ be a $\beta$-stable subordinator \cite{Applebaum:2009} with $0<\beta<1$ and characterized by its characteristic function $\mathbb{E}[\textrm{e}^{-ut(s)}]=\textrm{e}^{-su^\beta}$.
The corresponding inverse process, called inverse $\beta$-stable subordinator $s(t)$, has its characteristic function \cite{Mijena} $\mathbb{E}[\textrm{e}^{-\lambda s(t)}]=E_\beta(-\lambda t^\beta)$, with $E_{\beta}(t)=\sum_{k=0}^{\infty}\frac{t^k}{\Gamma(\beta k+1)}$ being a Mittag-Leffler function \cite{Erdelyi:1981}. So all moments of the inverse $\beta$-stable subordinator are finite. Besides that, the Laplace transform ($t\rightarrow u$) of the probability density function (pdf) $f(s,t)$ of the inverse $\beta$-stable subordinator $s(t)$ is \cite{Baule:2005}
\begin{equation}\label{Lfst}
  \mathcal{L}_{t\rightarrow u}[f(s,t)]=\int_0^\infty f(s,t) \textrm{e}^{-u t} \textrm{d}t= u^{\beta-1}\textrm{e}^{-su^\beta}.
\end{equation} 

In this paper, we mainly discuss some properties, such as, moments, diffusion type, covariance structure, of two kinds of stochastic processes subordinated by inverse $\beta$-stable process.
One is the time-changed non-Markovian Langevin system and another one the time-changed tempered fractional Brownian motion (tfBm).
In the first part, for long times, the time-changed tempered fractional Langevin equation (tfLe) could describe the subdiffusion for the case $0<\beta<\frac{1}{2}$ and the superdiffusion for the case $\frac{1}{2}<\beta<1$, even normal diffusion when $\beta=\frac{1}{2}$.
This is quite different from a common impression that the inverse $\beta$-stable subordinator ($0<\beta<1$) generally aims to yield a subdiffusion.
In the second part, for the time-changed tfBm by the inverse $\beta$-stable subordinator, its MSD and covariance function all tend to a constant at the rate $t^{-\beta}$, which is independent of the Hurst index $H$, while the ones of the original tfBm tend to a constant at the rate $t^{H-\frac{1}{2}}\e^{-\lambda t}$. All these results are verified by numerical simulations.

The structure of this paper is as follows. In section \ref{two}, the time-changed fractional Langevin equation (fLe) and tfLe with inverse $\beta$-stable subordinator are introduced. We discuss some properties of these subordinated processes, detecting the slower diffusion phenomenon than the original processes. Especially, the time-changed tfLe can display both subdiffusion and superdiffusion behaviors, depending on the value of $\beta$. Based on these observations, we discuss the differences between the time-changed tfLe and Langevin equation with biasing external force. In section \ref{three}, we introduce the time-changed tfBm by inverse $\beta$-stable process, discussing some properties such as the diffusion type and covariance structure, and making some comparisons between the subordinated tfBm and the original one. Finally, we conclude the paper with some remarks in section \ref{four}.

\section{Subordinated Langevin dynamics}\label{two}
In this section, we consider some time-changed non-Markovian Langevin systems (fLe and tfLe), subordinated by inverse $\beta$-stable process. The corresponding moments and diffusion types, as well as the Klein-Kramers equation and correlation functions are detailedly discussed. Then we compare two subordinated Langevin systems: tfLe and Langevin equation with biasing external potential. For long times, though their correlation functions and Klein-Kramers equations are completely different, the evolutions of their moments are found to be similar except the coefficients.

\subsection{Subordinated (fractional) Langevin equation}
The most basic Gaussian process, describing normal diffusion, is Brownian motion with its corresponding Langevin equation \cite{Coffey:04}
\begin{equation}\label{CLE1}
 \dot{x}(t)=\gamma(t),
\end{equation}
where $x(t)$ is the particle displacement, and the random fluctuation force $\gamma(t)$ is Gaussian white noise. Modify the physical time $t$ as the operational time $s$ and consider the coupled Langevin equation
\begin{equation}\label{CLE2}
 \dot{x}(s)=\gamma(s),~~~~  \dot{t}(s)=\eta(s),
\end{equation}
where the Gaussian white noise $\gamma(s)$ and the fully skewed $\beta$-stable L\'{e}vy noise $\eta(s)$ \cite{Schertzer:2001} are independent noise sources. So the random time transformation function $t(s)$ is a $\beta$-stable subordinator with $0<\beta<1$ as usual.
With the inverse $\beta$-stable subordinator $s(t)$, the combined process in physical time $t$ is $X(t):=x(s(t))$.
The coupled Langevin system (\ref{CLE2}) describing subdiffusion is the continuous realization of the CTRW models with power law distributed waiting time and normal distributed jump length \cite{Metzler:2000}, which has been proposed by Fogedby in \cite{Fogedby:1994}. Compared to (\ref{CLE1}), the subordinator $t(s)$ in (\ref{CLE2}) essentially changes the distribution of waiting time and thus eventually slows down the diffusion, i.e., turning normal diffusion into subdiffusion.

Fractional Langevin equation \cite{Deng:2009,Lutz:2001}, still describing Gaussian process, reads
\begin{equation}\label{fLe}
\dot{x}(t)=v(t),  \qquad  \dot{v}(t)=-\int_0^t(t-\tau)^{2H-2}v(\tau)\textrm{d}\tau+ \varrho\gamma(t).
\end{equation}
{The coefficient $\varrho$ is $[k_BT/(2D_HH(2H-1))]^{1/2}$ with the Hurst parameter $1/2<H<1$, the Boltzmann constant is $k_B$, absolute temperature is $T$ of the environment, and $D_H=[\Gamma(1-2H)\cos(H\pi)]/(2H \pi)$. The fractional Gaussian noise $\gamma(t)$ is a stationary Gaussian process with the mean $\langle\gamma(t)\rangle=0$ and the covariance
\begin{equation}
  \langle\gamma(t_1)\gamma(t_2)\rangle=2D_HH(2H-1)|t_1-t_2|^{2H-2}, \qquad  t_1,t_2>0.
\end{equation}
We assume that the initial velocity  $\dot x(0)=v_0$ satisfies thermal initial condition $v_0^2=k_BT$.
The first and second moments of the stochastic process $x(t)$ in (\ref{fLe}) are given in \cite{Deng:2009}}
\begin{eqnarray}\label{fLe_moments}
\langle x(t)\rangle=\sqrt{k_BT} \, E_{2H,2}(-\Gamma(2H-1)t^{2H})\,t, \nonumber\\
\langle x^2(t)\rangle=2k_BT \, E_{2H,3}(-\Gamma(2H-1)t^{2H})\,t^2,
\end{eqnarray}
with two-parameter Mittag-Leffler function $E_{\alpha,\beta}(t)=\Sigma_{n=1}^\infty \frac{t^n}{\Gamma(\alpha n+\beta)}$, which has the asymptotic expression $E_{\alpha,\beta}(-\gamma_\alpha t^\alpha)\simeq[\gamma_\alpha t^\alpha\Gamma(\beta-\alpha)]^{-1}$ for large $t$ and the Laplace transform $\mathcal{L}_{t\rightarrow u}[t^{\beta-1}E_{\alpha,\beta}(-\gamma_\alpha t^\alpha)]=u^{\alpha-\beta}(u^\alpha+\gamma_\alpha)^{-1}$ \cite{Podlubny:1999,Erdelyi:1954}.
Using the asymptotic expression of Mittag-Leffler function, the MSD of the trajectory sample $x(t)$ for large $t$ is
\begin{equation}\label{fLesub1}
  \langle (\Delta x(t))^2\rangle\simeq  \frac{2k_BT}{\Gamma(2H-1)\Gamma(3-2H)}\,t^{2-2H}.
\end{equation}
Since $1/2<H<1$, the Langevin system (\ref{fLe}) undergoes subdiffusion, which can model the dynamics of a single protein molecule \cite{Kou:2004}.
If the solution $x(t)$ of fLe (\ref{fLe}) is subordinated by an inverse $\beta$-stable subordinator $s(t)$ with $0<\beta<1$, then the subordinated stochastic process could be described by the following coupled fractional Langevin equation
\begin{eqnarray}\label{fLe_sub}
\fl \dot{x}(s)=v(s),  \qquad  \dot{v}(s)=-\int_0^s(s-\tau)^{2H-2}v(\tau)\textrm{d}\tau+ \varrho\gamma(s), \qquad \dot{t}(s)=\eta(s).
\end{eqnarray}
The pdf of the subordinated process $X(t):=x(s(t))$ can be written as \cite{Baule:2005,Barkai:2001}
\begin{equation}\label{pdf_subor}
p(x, t)=\int_0^\infty  p_0(x,s)f(s,t) \textrm{d}s,
\end{equation}
where $p_0(x,s)$ is {the} pdf of the original process $x(s)$ and $f(s,t)$ is {the} pdf of the inverse $\beta$-stable subordinator $s(t)$. The moments of subordinated process $X(t)$ could be obtained by the relation
\begin{equation}\label{relation}
\mathcal{L}_{t\rightarrow u}\langle X^n(t)\rangle=u^{\beta-1}\mathcal{L}_{s\rightarrow u^\beta}\langle x^n(s)\rangle
\end{equation}
in Laplace space, which could be got by multiplying $x^n$ on both sides of the equation (\ref{pdf_subor}) and integrating about $x$, {together with a} formula $\mathcal{L}_{t\rightarrow u}[f(s,t)]=u^{\beta-1}\textrm{e}^{-su^\beta}$. Then the first and second moments of $X(t)$ in subordinated fLe (\ref{fLe_sub}) can be obtained directly from (\ref{fLe_moments}) and (\ref{relation}) that
 \begin{eqnarray}
\langle X(t)\rangle=\sqrt{k_BT} \, E_{2H\beta,\beta+1}(-\Gamma(2H-1)t^{2H\beta})\,t^\beta, \\
\langle X^2(t)\rangle=2k_BT \, E_{2H\beta,2\beta+1}(-\Gamma(2H-1)t^{2H\beta})\,t^{2\beta},\nonumber
\end{eqnarray}
which are consistent with (\ref{fLe_moments}) in the case $\beta=1$. The MSD of subordinated process $X(t)$ for large physical time $t$ is
$$\langle (\Delta X(t))^2\rangle\simeq  \frac{2k_BT}{\Gamma(2H-1)\Gamma((2-2H)\beta+1)}\,t^{(2-2H)\beta}$$
with $0<\beta<1$. It also undergoes subdiffusion and  become slower than original process (\ref{fLesub1}).
It can be seen that after performing the  $\beta$-stable subordination on the fLe, the corresponding MSD can be easily obtained by replacing the parameter $H$ in the MSD of fLe with $1-(1-H)\beta$.  This simple way of obtaining the MSD for subordinated process does not hold for tfLe, which undergoes ballistic diffusion.



\subsection{Subordinated tempered fractional Langevin equation}
Tempered fractional Langevin equation is driven by tempered fractional Gaussian noise (tfGn) $\gamma(t)$, which has been detailedly discussed in \cite{Chen:2017}. It is also a Gaussian process and can be written as
\begin{equation}\label{tfLe}
\dot{x}(t)=v(t),  \qquad  \dot{v}(t)=- \int_0^tK(t-\tau)v(\tau)\textrm{d}\tau+\varrho\gamma(t),
\end{equation}
where $\varrho=\sqrt{2k_BT}$. The kernel $K(t)=2\langle\gamma(0)\gamma(t)\rangle=h^{-2}(C_{t+h}^2|t+h|^{2H}+C_{t-h}^2|t-h|^{2H}-2C_t^2|t|^{2H})$
for a sufficient small  $h$,  with  $0<H<1$ and
\begin{equation}\label{C_t^2}
C_t^2=\frac{2\Gamma(2H)}{(2\lambda|t|)^{2H}}-\frac{2\Gamma(H+\frac{1}{2})K_H(\lambda|t|)}{\sqrt{\pi}(2\lambda|t|)^H},
\end{equation}
where $K_H(t)$ is the modified Bessel function of second kind. The initial velocity satisfies thermal initial condition, i.e., $v_0^2=k_BT$. 
For fixed small $\lambda$, with the time evolution the first and second moments of particle trajectory $x(t)$ behave like
\begin{equation}\label{1}
\langle x(t) \rangle: ~ \sqrt{k_BT}t\rightarrow Bt^{1-2H}\rightarrow At
\end{equation}
and
\begin{equation}\label{2}
\langle x^2(t) \rangle: ~ D t^{2+2H}+k_BT t^2\rightarrow Ct^{2-2H}\rightarrow \sqrt{k_BT}At^2.
\end{equation}
Here $A=\sqrt{k_BT}/[1+2\Gamma(2H)(2\lambda)^{-2H}],\, B=\sqrt{k_BT}/[2D_H\Gamma^2(H+1/2)\Gamma(2H+1)\Gamma(2-2H)],\, C=k_BT/[D_H\Gamma^2(H+1/2)\Gamma(2H+1)\Gamma(3-2H)]$, and  $D=4D_H\Gamma^2(H+1/2)k_BT/(H+1)$.
In particular, for the short time, from (\ref{2}) it can be seen that $D t^{2+2H}$ dominates MSD. While for long times, the particle displays ballistic diffusion, a special superdiffusion. 
Another model displaying ballistic diffusion is the celebrated L\'{e}vy walk \cite{Zaburdaev:2015}, where the particle moves with a constant speed and only changes its direction at a random time, and the waiting time obeys power law distribution with the exponent less than 1.
One obvious difference is that the process $x(t)$ described by tfLe in (\ref{tfLe})  is a Gaussian process while the L\'{e}vy walk model is not. The connection between L\'{e}vy walk model and the corresponding coupled Langevin system with $\beta$-stable subordinator is presented in \cite{Eule:2012}.

\begin{figure}[!htb]
\flushright
\begin{minipage}{0.8\linewidth}
  \centerline{\includegraphics[width=10cm,height=6.5cm]{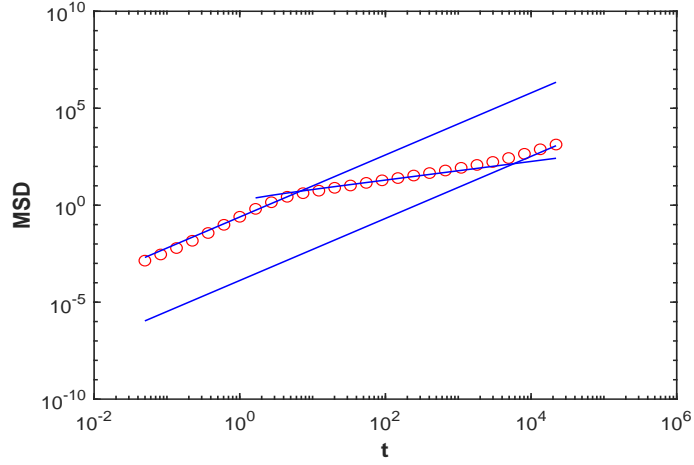}}
\end{minipage}
  \caption{Evolution of the MSD of the subordinated tfLe.  Solid lines are the analytical results (\ref{tfLe_MSD}) and the circle-markers are the computer simulations with the physical time $T=2\times10^4$. Parameter values: $H=0.7$, $\lambda=0.001$, $\beta=0.8$, and $k_BT=1$.  }\label{MSD}
\end{figure}

Now, we turn to the time-changed tfLe coupled with $\beta$-stable subordinator
\begin{eqnarray}\label{sub_tfLe}
\fl \dot{x}(s)=v(s),  \qquad  \dot{v}(s)=- \int_0^sK(s-\tau)v(\tau)\textrm{d}\tau+\varrho\gamma(s),\qquad \dot{t}(s)=\eta(s).
\end{eqnarray}
According to (\ref{relation}), with the time evolution the first and second moments of the subordinated process $X(t):=x(s(t))$ behave as 
\begin{equation}\label{3}
\langle X(t) \rangle: ~ \frac{\sqrt{k_BT}}{\beta\Gamma(\beta)}\,t^\beta \rightarrow E\,t^{(1-2H)\beta}\rightarrow \frac{A}{\beta\Gamma(\beta)}\,t^\beta
\end{equation}
and
\begin{equation}\label{4}
\langle X^2(t) \rangle: ~ \frac{k_BT}{\beta\Gamma(2\beta)}\,t^{2\beta} \rightarrow F\,t^{(2-2H)\beta}\rightarrow \frac{\sqrt{k_BT}A}{\beta\Gamma(2\beta)}\,t^{2\beta},
\end{equation}
where $E=\sqrt{k_BT}/[2D_H\Gamma^2(H+1/2)\Gamma(2H+1)\Gamma((1-2H)\beta+1)]$ and $ F=k_BT/[D_H\Gamma^2(H+1/2)\Gamma(2H+1)\Gamma((2-2H)\beta+1)]$. These asymptotic behaviors are consistent with (\ref{1}) and (\ref{2}) when $\beta=1$. With the time evolution, the MSD of this subordinated tfLe goes like
\begin{eqnarray}\label{tfLe_MSD}
\langle (\Delta X(t))^2 \rangle: &   \left(\frac{k_BT}{\beta\Gamma(2\beta)}-\frac{k_BT}{\beta^2\Gamma^2(\beta)}\right)  t^{2\beta}\rightarrow Ft^{(2-2H)\beta}-E^2t^{2(1-2H)\beta}  \\
  &~\rightarrow  \left(\frac{\sqrt{k_BT}A}{\beta\Gamma(2\beta)}-\frac{A^2}{(\beta\Gamma(\beta))^2}\right)t^{2\beta}\nonumber.
\end{eqnarray}
The simulation results of MSD are given in figure \ref{MSD}, displaying the transition procedure with the time evolution.  
To observe the middle stage clearly, we take a moderately small $\lambda=0.001$. In figure \ref{MSD}, it can be found that the simulation results of MSD are consistent with the theoretical ones (\ref{tfLe_MSD}) through the whole procedure. Especially, for large times, the diffusion of particle described by the subordinated tfLe (\ref{sub_tfLe}) is slower than the original process exhibiting ballistic diffusion, and could be subdiffusion when $0<\beta<1/2$, superdiffusion when $1/2<\beta<1$, and even normal diffusion as $\beta=1/2$.

\begin{figure}[!htb]
\flushright
\begin{minipage}{0.65\linewidth}
  \centerline{\includegraphics[scale=0.45]{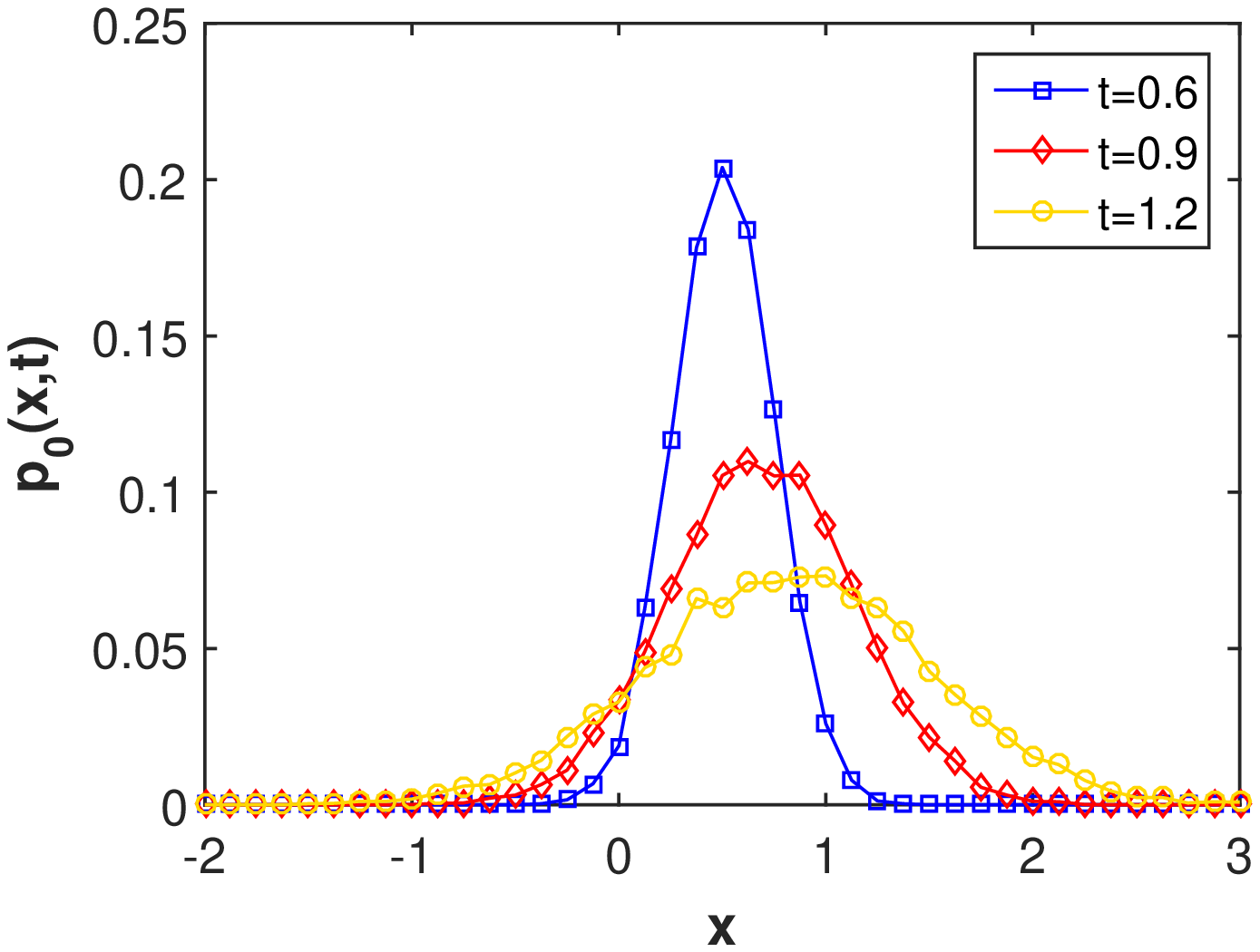}}
  \centerline{(a)}
\end{minipage}
\hfill
\begin{minipage}{0.31\linewidth}
  \centerline{\includegraphics[scale=0.45]{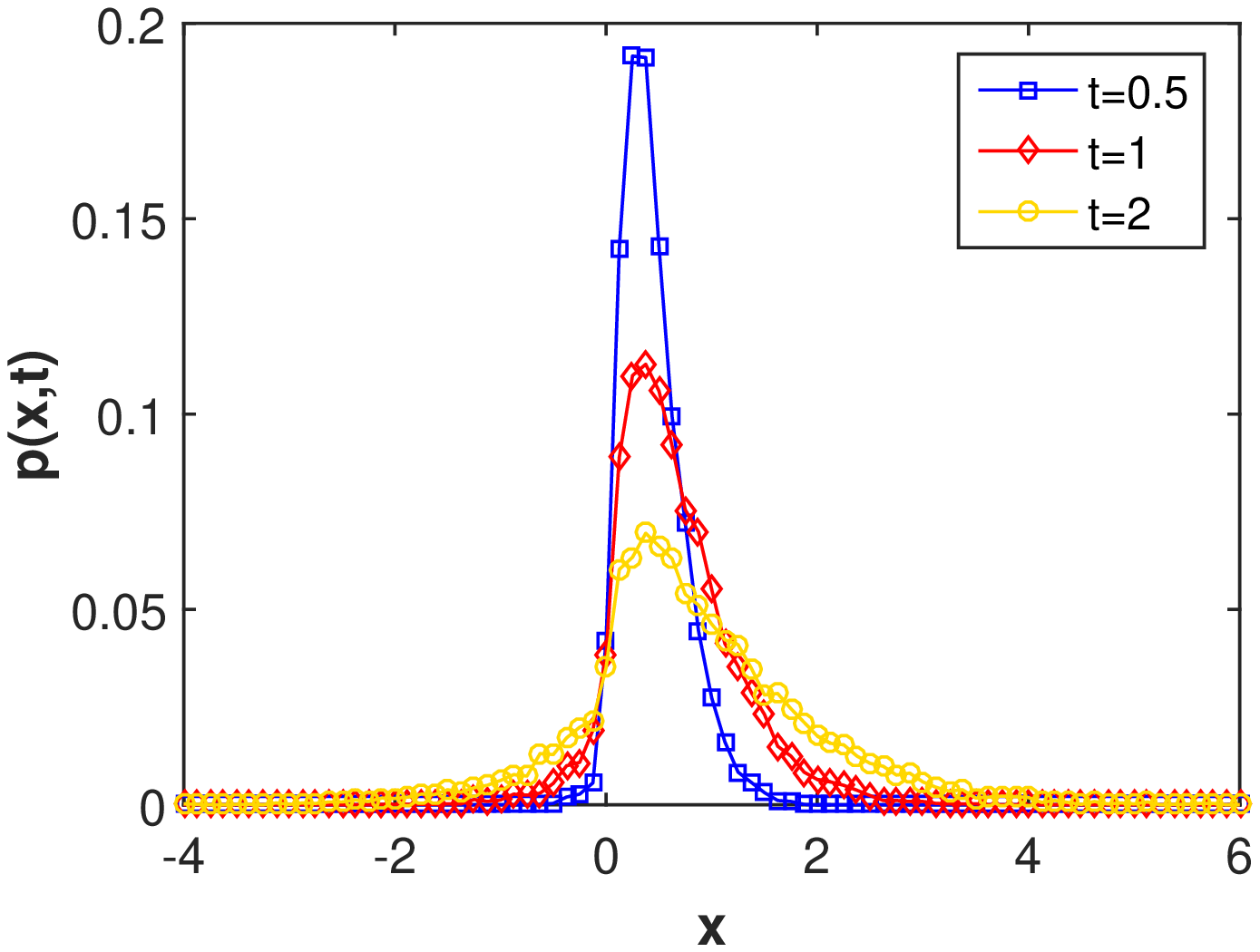}}
  \centerline{(b)}
\end{minipage}
  \caption{Time evolution of the pdf of tfLe for different times $t=0.6, 0.9,$ and $1.2$ displayed in (a) and pdf of subordinated tfLe for different times $t=0.5,\, 1,$ and $2$ shown in (b). Parameter values: $H=0.7$, $\lambda=0.1$, $\beta=0.8$, and $k_BT=1$; and the number of simulation trajectories is $8000$.}\label{PDF}
\end{figure}

The simulation results of the pdf $p_0(x,t)$ of tfLe and the pdf $p(x,t)$ of subordinated tfLe for different times $t$ are shown in figure {\ref{PDF}} and we can find that the subordinated process $X(t)$ is no longer Gaussian process while the original process $x(t)$ is. Non-zero mean of the original process $x(t)$ results in a right shift of the peak of the symmetric pdf curve in $(a)$, while the  non-zero mean of the subordinated process $X(t)$ leads to an asymmetry of the pdf curve in $(b)$. The asymmetry pdf curve is similar to figure $1$ in \cite{Eule:2009} that a biasing external force, which contributes to the non-zero mean acts only at the time of the jumps but not affects the dynamics of the diffusing particle during the waiting periods. In the next subsection, we will detailedly make a comparison between the subordinated tfLe and the Langevin equation with biasing external force.

From the subordinated tfLe (\ref{sub_tfLe}), one can derive the corresponding Klein-Kramers equation, which governs the joint pdf $p(x,v,t)$ of finding the particle at time $t$ and position $x$ with velocity $v$. In general, for two-dimensional Gaussian processes $y_i(t),i=1,2$, denoting $\Delta y_i(t):=y_i(t)-\langle y_i(t)\rangle$, their joint pdf is
\begin{equation}\label{joint_pdf}
p(y_1,y_2,t)=\frac{1}{2\pi \sqrt{|R|}}\textrm{exp}\left[-\frac{1}{2}(\Delta y)^{T}R^{-1}\Delta y\right],
\end{equation}
with
\begin{equation*}
  \Delta y=\left(\begin{array}{c}
    \Delta y_1(t)\\ \Delta y_2(t)
  \end{array}\right)
  ,\qquad
R=
\left(\begin{array}{cc} \langle (\Delta y_1(t))^2\rangle & \langle \Delta y_1(t)\Delta y_2(t)\rangle \\ \langle \Delta y_2(t)\Delta y_1(t)\rangle & \langle (\Delta y_2(t))^2\rangle \end{array} \right),
\end{equation*}
and $(\Delta y)^{T}$ denotes the transposition of $\Delta y$. Taking Fourier transform ($y_1\rightarrow k_1$, $y_2\rightarrow k_2$) about (\ref{joint_pdf}), one gets that
\begin{eqnarray*}
\fl p(k_1,k_2,t)
 =\int_{-\infty}^{+\infty}\int_{-\infty}^{+\infty}p(y_1,y_2,t)\e^{-ik_1y_1-ik_2y_2}\textrm{d}y_1\textrm{d}y_2  \\
\fl =\textrm{exp}\left[{-ik_1\langle y_1(t)\rangle-ik_2\langle y_2(t)\rangle}
   {-\frac{k_1^2}{2}\langle (\Delta y_1(t))^2\rangle-\frac{k_2^2}{2}\langle (\Delta y_2(t))^2\rangle}
   {-k_1k_2\langle \Delta y_1(t)\Delta y_2(t)\rangle}\right].
\end{eqnarray*}
Then taking partial derivative w.r.t. $t$, and performing inverse Fourier transform, one gets the equation of the joint pdf $p(y_1,y_2,t)$, namely,
\begin{eqnarray}
\fl\frac{\partial p(y_1,y_2,t)}{\partial t}=
&-\frac{\textrm{d}\langle y_1(t)\rangle}{\textrm{d}t} \cdot\frac{\partial p(y_1,y_2,t)}{\partial y_1}
-\frac{\textrm{d}\langle y_2(t)\rangle}{\textrm{d}t} \cdot\frac{\partial p(y_1,y_2,t)}{\partial y_2}\nonumber\\
\fl&+\frac{1}{2}\frac{\textrm{d}\langle (\Delta y_1(t))^2\rangle}{\textrm{d}t} \cdot\frac{\partial^2 p(y_1,y_2,t)}{\partial y_1^2}
+\frac{1}{2}\frac{\textrm{d}\langle (\Delta y_2(t))^2\rangle}{\textrm{d}t} \cdot\frac{\partial^2 p(y_1,y_2,t)}{\partial y_2^2}\nonumber\\
\fl&+\frac{\textrm{d}\langle\Delta y_1(t)\Delta y_2(t)\rangle}{\textrm{d}t} \cdot\frac{\partial^2 p(y_1,y_2,t)}{\partial y_1 \partial y_2}.
\end{eqnarray}
For tfLe (\ref{tfLe}), we know that the trajectory sample $x(t)$ and the corresponding velocity $v(t)$ all obey normal distribution. Considering the large time case in (\ref{1})-(\ref{2}) and the results $\langle[x(t)-\langle x(t)\rangle][v(t)-\langle v(t)\rangle]\rangle\simeq \overline{A}t$, $\langle v(t)\rangle\simeq A$, $\langle [v(t)-\langle v(t)\rangle]^2\rangle\simeq 2\overline{A}$ with $\overline{A}=2\Gamma(2H)(2\lambda)^{-2H}A^2$ in \cite{Chen:2017}, the Klein-Kramers equation of tfLe for large time $t$ can be represented as
\begin{eqnarray}\label{FP_tfLe}
\frac{\partial p_0(x,v,t)}{\partial t}=-A\frac{\partial p_0(x,v,t)}{\partial x}
+\overline{A}t\frac{\partial^2 p_0(x,v,t)}{\partial x^2}+\overline{A}\frac{\partial^2 p_0(x,v,t)}{\partial x \partial v}.
\end{eqnarray}

\begin{figure}[!htb]
\flushright
\begin{minipage}{0.8\linewidth}
  \centerline{\includegraphics[width=10cm,height=6.5cm]{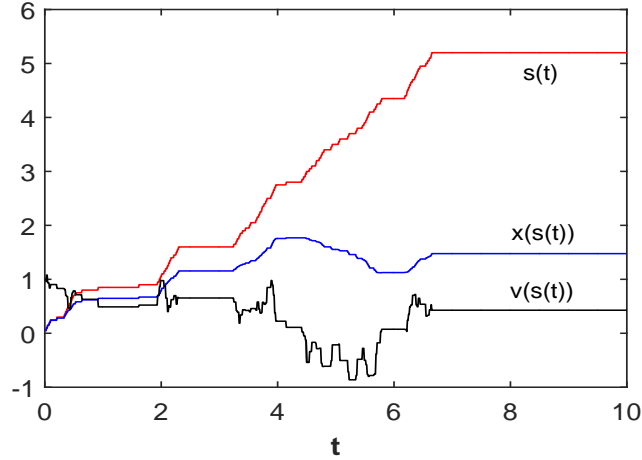}}
  \end{minipage}
\caption{Sample trajectories of inverse $\beta$-stable subordinator $s(t)$, position process $x(s(t))$, and velocity process $v(s(t))$.  Parameter values: $H=0.7$, $\lambda=0.1$, $\beta=0.8$, and $k_BT=1$.}\label{x_v}
\end{figure}

Applying the method in \cite{Gajda:2011,Hahn:2011},  the joint pdf $p(x,v,t)$ of the subordinated process $[X(t),V(t)]:=[x(s(t)),v(s(t))]$ described by the tfLe coupled with $\beta$-stable subordinator in model (\ref{sub_tfLe}), has the form
\begin{equation}\label{P-P0}
p(x,v, t)=\int_0^\infty  p_0(x,v,s)f(s,t) \textrm{d}s,
\end{equation}
where $p_0(x,v,s)$ is the joint pdf of the original process, i.e., the solution of (\ref{FP_tfLe}) by replacing $t$ with $s$.
The Laplace transform ($t\rightarrow u$) in (\ref{P-P0}) gives the equality in Laplace space:
\begin{equation}\label{P-P02}
p(x,v,u)=u^{\beta-1}p_0(x,v,u^{\beta}).
\end{equation}
Combining it with the equation (\ref{FP_tfLe}) in Laplace space gives
\begin{eqnarray*}
\fl up(x,v,u)-p(x,v,0)=
&-Au^{1-\beta}\frac{\partial p(x,v,u)}{\partial x}
-\frac{\overline{A}(1-\beta)}{\beta}u^{1-2\beta}\frac{\partial^2 p(x,v,u)}{\partial x^2}  \\
\fl&-\frac{\overline{A}}{\beta}u^{2-2\beta}\frac{\partial}{\partial u}\frac{\partial^2 p(x,v,u)}{\partial x^2}
+\overline{A}u^{1-\beta}\frac{\partial^2 p(x,v,u)}{\partial x \partial v}.
\end{eqnarray*}
Then taking inverse Laplace transform, the Klein-Kramers equation of the subordinated process $[X(t),V(t)]$ for large time $t$ is
\begin{eqnarray}\label{sub_P}
\frac{\partial p(x,v,t)}{\partial t}=
&~\frac{\overline{A}}{\beta}\Big[D_t^{2-2\beta}t-(1-\beta)D_t^{1-2\beta}\Big]\frac{\partial^2 p(x,v,t)}{\partial x^2}\nonumber\\
&-A D_t^{1-\beta}\frac{\partial p(x,v,t)}{\partial x}+\overline{A} D_t^{1-\beta}\frac{\partial^2 p(x,v,t)}{\partial x \partial v},
\end{eqnarray}
where $D_t^\ast$ is the Riemann-Liouville fractional derivative \cite{Podlubny:1999}, defined as
\begin{eqnarray}
D_t^qW(t)=\frac{1}{\Gamma(m-q)}\frac{\partial^{m}}{\partial t^{m}}\int_0^t (t-t')^{m-1-q}W(t')\textrm{d}t',
\end{eqnarray}
 for $m-1< q \leq m$ and its Laplace transform
\begin{eqnarray}
\mathcal{L}_{t\rightarrow u}[D_t^qW(t)]=u^qW(u)-\sum_{k=0}^{m-1}u^k[D_t^{q-k-1}W(t)]_{t=0}.
\end{eqnarray}
 The Klein-Kramers equation (\ref{sub_P}) is consistent with (\ref{FP_tfLe}) when $\beta=1$. Integrating both sides of this equation about $v$ yields the Fokker-Planck equation for the marginal pdf $p(x,t)$ of position of the particle
\begin{eqnarray}\label{PP}
\frac{\partial p(x,t)}{\partial t}=-A D_t^{1-\beta}\frac{\partial p(x,t)}{\partial x}
+\frac{\overline{A}}{\beta}\Big[D_t^{2-2\beta}t-(1-\beta)D_t^{1-2\beta}\Big]\frac{\partial^2 p(x,t)}{\partial x^2}.
\end{eqnarray}
We simulate the position process $X(t)=x(s(t))$ and velocity process $V(t)=v(s(t))$ in figure \ref{x_v}.
It can be seen that the constant time periods of $s(t)$ (red curve) represent the trapping events, where $X(t)$ (blue curve) and $V(t)$ (black curve) are keeping their current states respectively. And then when the trapping event is finished, the particle is released with the same velocity as prior.
Note that $X(t)$ and $V(t)$ no longer satisfy the Newtonian relation but $\frac{\textrm{d}}{\textrm{d}t}\langle X(t)\rangle=D_t^{1-\beta}\langle V(t)\rangle$ due to the additional waiting time average \cite{Gajda:2011,Metzler:00}.

Since tfLe (\ref{tfLe}) describes Gaussian process, all moments exist and can be obtained by the formula
\begin{equation}
\langle x^n(t)\rangle=  \mathcal{L}^{-1}_{u\rightarrow t}\left[ i^n\left.\frac{\partial^np_0(k,u)}{\partial k^n}\right|_{k=0} \right].
\end{equation}
So all moments of the subordinated process $X(t)$ exist; for large time $t$, by the formula (\ref{relation}), there are
\begin{equation}\label{moments}
\fl \langle X(t)\rangle\simeq \frac{A}{\beta\Gamma(\beta)}\,t^\beta,
\quad \langle X^2(t)\rangle\simeq \frac{\sqrt{k_BT}A}{\beta\Gamma(2\beta)}\,t^{2\beta},
\quad \langle X^3(t)\rangle\simeq G\,t^{3\beta},~...,
\end{equation}
where $G=2A^3(1+6\Gamma(2H)(2\lambda)^{-2H})/[\beta\Gamma(3\beta)]$. For the case of $\beta=1$, these moments go back to the ones of original process $x(t)$.

It is not enough to characterize a stochastic process solely by its pdf and moments, the correlation function of this stochastic process is also needed to characterize the correlation of different times. Baule and Friedrich \cite{Baule:2005} derive the two-time pdf $h(s_2,t_2;s_1,t_1)$ of the inverse $\beta$-stable process $s(t)$ in Laplace space ($t_1\rightarrow u_1,t_2\rightarrow u_2$):
\begin{eqnarray}\label{twotimepdf}
\fl h(s_2,u_2;s_1,u_1)=&\delta(s_2-s_1)\,\frac{u_1^\beta-(u_1+u_2)^\beta+u_2^\beta}{u_1u_2}  \,\textrm{e}^{-s_1(u_1+u_2)^\beta}\nonumber\\
&+\Theta (s_2-s_1)  \,\frac{u_2^\beta[(u_1+u_2)^\beta-u_2^\beta]}{u_1u_2}  \,\textrm{e}^{-s_1(u_1+u_2)^\beta}\textrm{e}^{-(s_2-s_1)u_2^\beta}\nonumber\\
&+\Theta (s_1-s_2)  \,\frac{u_1^\beta[(u_1+u_2)^\beta-u_1^\beta]}{u_1u_2}  \,\textrm{e}^{-s_2(u_1+u_2)^\beta}\textrm{e}^{-(s_1-s_2)u_1^\beta},
\end{eqnarray}
where $\Theta(x)$ is the Heaviside step function and $\Theta(x)=1$ for $x>0$, $\Theta(x)=0$ for $x<0$, $\Theta(x=0)=1/2$. Using $\langle X(t_2)X(t_1)\rangle=\int_0^\infty\int_0^\infty h(s_2,t_2;s_1,t_1)\langle x(s_2)x(s_1)\rangle \textrm{d}s_1 \textrm{d}s_2 $ and (\ref{twotimepdf}), as well as the autocorrelation function $\langle x(s_1)x(s_2)\rangle$ of particle trajectory $x(s)$ of tfLe (\ref{tfLe}) for large $s_1$ and $s_2$: $\langle x(s_1)x(s_2)\rangle\simeq\sqrt{k_BT}As_1s_2$ in \cite{Chen:2017}, one could obtain the covariance function $\textrm{cov}[X(t_1),X(t_2)]$ of the subordinated process $X(t)=x(s(t))$ described by (\ref{sub_tfLe}) for fixed large $t_2$ and extremely large $t_1$ ($t_1>t_2$):
\begin{eqnarray}\label{cov}
\fl\textrm{cov}[X(t_1),X(t_2)]
&=\langle X(t_1)X(t_2)\rangle-\langle X(t_1)\rangle\langle X(t_2)\rangle\nonumber\\
\fl&\simeq\sqrt{k_BT}A\left[\frac{t_2^{2\beta}}{2\beta\Gamma(2\beta)}
+\frac{t_1^{2\beta}}{\beta\Gamma^2(\beta)}B\left(\beta,\beta+1;\frac{t_2}{t_1}\right)\right]
-\frac{A^2}{\beta^2\Gamma^2(\beta)}t_1^\beta t_2^\beta\nonumber\\
\fl&\simeq \frac{\sqrt{k_BT}A}{2\beta\Gamma(2\beta)}\,t_2^{2\beta}+
\frac{2\Gamma(2H)(2\lambda)^{-2H}A^2}{\beta^2\Gamma^2(\beta)}\,t_2^\beta t_1^\beta,
\end{eqnarray}
where $B(a,b;z)=\int_0^z\tau^{a-1}(1-\tau)^{b-1}\textrm{d}\tau$ is the incomplete Beta function \cite{Abramowitz:1972} and
$$B\left(\beta,\beta+1;\frac{t_2}{t_1}\right)\simeq \frac{1}{\beta}\left(\frac{t_2}{t_1}\right)^{\beta}$$
for fixed $t_2$ and large $t_1$.
Then the correlation function $\textrm{corr}[X(t_1),X(t_2)]$ of the subordinated process $X(t)$ is
\begin{eqnarray}\label{co1}
\fl\textrm{corr}[X(t_1),X(t_2)]&=\frac{\textrm{cov}[X(t_1),X(t_2)]}{\sqrt{\langle[X(t_1)-\langle X(t_1)\rangle]^2\rangle\langle[X(t_2)-\langle X(t_2)\rangle]^2\rangle}}\nonumber\\
&\simeq \frac{\sqrt{k_BT}\beta\Gamma^2(\beta)}{2\sqrt{k_BT}\beta\Gamma^2(\beta)-2A\Gamma(2\beta)}t_2^\beta t_1^{-\beta}+\frac{2\Gamma(2H)(2\lambda)^{-2H}A^2\Gamma(2\beta)}{\sqrt{k_BT}A\beta\Gamma^2(\beta)-A^2\Gamma(2\beta)}.
\end{eqnarray}
The case of $\beta=1$ is that the correlation function of tfLe $\textrm{corr}[x(t_1),x(t_2)]\simeq1$ for the limit $t_1\rightarrow \infty$. That is to say, with the lengthening of the time interval, the degree of correlation of $X(t_1)$ (or $x(t_1)$) and $X(t_2)$ (or $x(t_2)$) remains unchanged.

\subsection{Subordinated Langevin equation with biasing external force fields}
One important result in the previous subsection shows that $\beta$-stable subordinator slows down the original diffusion, regardless of whether it is subdiffusion, normal diffusion or superdiffusion. Especially, the effect of $\beta$-stable subordinator on the ballistic diffusion described by tfLe could produce different types of diffusion, depending on the value of $\beta$.
Eule \emph{et al} \cite{Eule:2012} show that the effect of subordination on normal diffusion is not limited to subdiffusion but can also produce superdiffusion.
In that paper, a Langevin system is  subordinated by an inverse $\beta$-stable subordinator $s(t)$, and the sample $x(t)=\int_0^t v(s(t')) \textrm{d}t'$ with normal distributed velocity $v(s)$ in operation time $s$  transforms from superdiffusion for short times to ballistic diffusion for long times.

Besides above, another common model is the coupled Langevin system \cite{Fogedby:1994,Cairoli:2015,Eule:2009,Denisov:2009}
 \begin{eqnarray}\label{LE_subor}
 \dot{x}(s)=F(x(s))+\sigma(x(s))\gamma(s),~~~~  \dot{t}(s)=\eta(s),
\end{eqnarray}
where $F(x)$ is a force field, $\sigma(x)$ is a multiplicative noise term, $\gamma(s)$ is Gaussian white noise, and $\eta(s)$ is a fully skewed $\beta$-stable L\'{e}vy noise with $0<\beta<1$. The external force field in this Langevin system is biased \cite{Eule:2009}, which means that the force acts as a bias only at the moment of an actual jump. It is essentially different from the decoupled external force field \cite{Eule:2009}, where the particle is affected by the external force during the whole waiting time period and the diffusion process is decoupled from the effect of force field.  The corresponding Fokker-Planck equation of $X(t):=x(s(t))$ in (\ref{LE_subor}) is
\begin{eqnarray}\label{P}
\frac{\partial p(x,t)}{\partial t}=\mathcal{L}_{FP}D_t^{1-\beta}p(x,t),
\end{eqnarray}
with the Fokker-Planck operator $\mathcal{L}_{FP}=-\frac{\partial}{\partial x} F(x)+\frac{1}{2}\frac{\partial^2}{\partial x^2} \sigma^2(x)$.
For the case $\beta=1$, it becomes the standard Fokker-Planck equation without temporal fractional operator $D_t^{1-\beta}$.

The equation (\ref{P}) can be commonly derived by three methods. The first one is based on the relation between the pdf of subordinated process and original
process that $p(x,u)=u^{\beta-1}p_0(x,u^\beta)$ in Laplace space similar to (\ref{P-P02}). The second one makes use of the It\^{o} formula in \cite{Cairoli:2015} by taking $p=0$ there. As for the last method, (\ref{P}) can be derived using the governing equation
\begin{equation}
  p(k,u)=\frac{1-w(u)}{u}\frac{1}{1-\psi(k,u)}
\end{equation}
in CTRW models \cite{Metzler:2000}, where $p(k,u)$ is the Fourier-Laplace transform ($x\rightarrow k, t\rightarrow u$) of $p(x,t)$, $w(u)$ is the Laplace transform of waiting time pdf $w(t)$, and $\psi(k,u)$ corresponds to the jump pdf $\psi(x,t)$.
Assuming $w(u)\simeq 1-(u\tau)^\beta$ as $u\rightarrow0$ with $0<\beta<1$ and $\psi(x,t)=\psi_0(x-v\tau_a,t)$, where $\tau_a$ is a microscopic advection time and $\psi_0(x,t)$ is the jump pdf in the case without external force, with form $\psi_0(k,u)=\frac{1}{1+(u\tau)^\beta}\textrm{e}^{-\rho^2k^2}$, we can obtain (\ref{P}) with some specified $F(x)$ and $\sigma^2(x)$ \cite{Metzler:2000}. The equation (\ref{P}) is the Galilei variant fractional diffusion-advection equation \cite{Compte:1997,Compte:97} since $p(x,t)\neq p_{v=0}(x-vt,t)$, where $ p_{v=0}(x,t)$ denotes the free propagator \cite{Metzler:2000}. In particular, the jump pdf $\psi(x,t)=\psi_0(x-v\tau_a,t)$ indicates that the external force field is the biasing force since $v\tau_a$ means that the force only acts at the moment of jump. But if adopting $\psi(x,t)=\psi_0(x-vt,t)$, the Galilei invariant fractional diffusion-advection equation \cite{Metzler:2000} can be obtained as
\begin{eqnarray}
\frac{\partial p(x,t)}{\partial t}+v\frac{\partial p(x,t)}{\partial x}=D_t^{1-\beta}K_\beta \frac{\partial ^2 p(x,t)}{\partial x^2}.
\end{eqnarray}
The external force here is decoupled and the MSD of this case still behaves as $\langle (\Delta x(t))^2\rangle\simeq\frac{2K_\beta}{\Gamma(1+\beta)}t^\beta$ with $0<\beta<1$, describing subdiffusion, being the same as the MSD of free particle with $F=0$.

Next, we pay attention to the moments of the coupled Langevin system (\ref{LE_subor}). Taking the constant force $F(x)=A$ and $\sigma^2(x)=\overline{A}$, one can obtain that the external process $x(s)$ obeys normal distribution $N(As,\overline{A}s)$ over operation time $s$. Then by formula (\ref{relation}), the moments of subordinated process $X(t)$ described by coupled Langevin system (\ref{LE_subor}) are as follows
\begin{eqnarray}\label{momentsnew}
\langle X(t)\rangle= \frac{A}{\beta\Gamma(\beta)}\,t^\beta,
\qquad \langle X^2(t)\rangle= \frac{A^2}{\beta\Gamma(2\beta)}\,t^{2\beta}+\frac{\overline{A}}{\beta\Gamma(\beta)}\,t^\beta,\nonumber\\
 \langle X^3(t)\rangle=\frac{2A^3}{\beta\Gamma(3\beta)}\,t^{3\beta}+\frac{3\overline{A}A}{\beta\Gamma(2\beta)}\,t^{2\beta},\qquad \cdots \,.
\end{eqnarray}
For long times, the asymptotic behavior of the moments are similar to the ones of subordinated tfLe (\ref{moments}) except the smaller coefficients. When $\beta=1$, all the above moments (\ref{momentsnew}) reduce to the ones of original process. The subordinated process $X(t)$ in (\ref{LE_subor}) is no longer Gaussian ($\beta\neq1$) and its MSD  is
\begin{equation*}
\langle (\Delta X(t))^2\rangle\simeq
\left\{
  \begin{array}{ll}
     \overline{A}t,~~~& \hbox{for}~~\beta=1, \\
     \left( \frac{A^2}{\beta\Gamma(2\beta)}-\frac{A^2}{\beta^2\Gamma^2(\beta)}\right)t^{2\beta},~~~& \hbox{for}~~\beta\neq 1,
  \end{array}
\right.
\end{equation*}
which is consistent with \cite{Compte:97} based on CTRW models.
It can be seen that the coupled Langevin system (\ref{LE_subor}) shows subdiffusion when $0<\beta<\frac{1}{2}$, superdiffusion when $\frac{1}{2}<\beta<1$, and normal diffusion when $\beta=1$ or $\frac{1}{2}$. More or less, it is a strange phenomenon that infinite mean waiting time produces superdiffusion. Compte \emph{et al} explain this paradox in \cite{Compte:97} that some stagnated particles are not continuously dragged by the stream and thus slow down the advancement of the center of mass of the particles, instead the main dispersion mechanism should be  convection. 
The MSD of (\ref{LE_subor}) is similar to the MSD $\langle (\Delta X(t))^2\rangle\simeq[\sqrt{k_BT}A/(\beta\Gamma(2\beta))-A^2/(\beta^2\Gamma^2(\beta))]t^{2\beta}$ with $0<\beta<1$ of subordinated tfLe in (\ref{tfLe_MSD}), except the smaller coefficient. However, the main dispersion mechanism of the subordinated Langevin equation with biasing external force (\ref{LE_subor}) is not produced by diffusion but convection, while the main dispersion mechanism of the subordinated tfLe (\ref{sub_tfLe}) is produced by diffusion itself.

From (\ref{LE_subor}), one can obtain the covariance function of the external process $x(s)$ as $\langle x(s_1)x(s_2)\rangle=A^2s_1s_2+\overline{A} \textrm {min}(s_1,s_2)$. For fixed $s_2$ and large $s_1$, the correlation function of $x(s)$ is
\begin{equation}
  \textrm{corr}[x(s_1),x(s_2)]\simeq s_2^{1/2}s_1^{-1/2},
\end{equation}
which means that the process $x(s)$ is long-range dependent  \cite{Wylomanska:2016}. Then the correlation function of the subordinated process $X(t)=x(s(t))$ can be obtained by (\ref{twotimepdf}) and $\langle X(t_2)X(t_1)\rangle=\int_0^\infty\int_0^\infty h(s_2,t_2;s_1,t_1)\langle x(s_2)x(s_1)\rangle \textrm{d}s_1 \textrm{d}s_2 $. So we get
\begin{equation}\label{co2}
  \textrm{corr}[X(t_1),X(t_2)]\simeq C(t_2)t_1^{-\beta}
\end{equation}
for fixed $t_2$ and large $t_1$, where $\beta\neq1$ and $C(t_2)$ is a constant depending on $t_2$. It indicates that the subordinated process $X(t)$ is also long-range dependent.

Comparing the above two coupled Langevin systems, i.e., the  subordinated tfLe (\ref{sub_tfLe}) and subordinated Langevin equation with biasing external force (\ref{LE_subor}), we find that for long times, the original processes and the subordinated ones have similar moments;  see (\ref{moments}) and (\ref{momentsnew}), respectively. Paying special attention to the MSDs of these two Langevin systems, although the original processes have different diffusion types (one is ballistic diffusion, another one is normal diffusion), the MSDs of the subordinated Langevin systems are similar, both being $t^{2\beta}$.
However, one mainly stems from the slow diffusion caused by the additional waiting time, while another one is because of the effect of biasing external force $F(x)$. In addition, the Fokker-Planck equations of the pdf $p(x,t)$ of the two subordinated processes are completely different (see (\ref{PP}) and (\ref{P})), although both have temporal fractional derivative and when $\beta=1$ both Fokker-Planck equations  reduce to the original one. Besides that, because of the differences of the noises (one is tfGn and another one is Gaussian white noise) and the complexity of the systems, the correlation structures of these two subordinated processes are quite different; see (\ref{co1}) and (\ref{co2}).

\section{Subordinated tempered fractional Brownian motion}\label{three}
From the above discussions, for subordinated Brownian motion (Bm),  the method of subordination can make the
jumps of Bm occur after long waiting times, which eventually slows down the diffusion and turns the original normal diffusion into subdiffusion. For time-changed fractional Brownian motion (fBm) by different subordinators, there have been many  literatures \cite{Hahn:2011,Wylomanska:2016,Mijena,Kumar:2017} presenting some of its properties, such as the covariance structure, ergodic property, and so on. Here we pay attention to the time-changed time fractional Brownian motion (tfBm) by inverse $\beta$-stable process, discussing its moments, covariance function, and the covariance function of its increments. We also compare these statistical quantities between the time-changed tfBm and the original one.

Tempered fractional Brownian motion is introduced by Meerschaert and Sabzikar \cite{Meerschaert:2013}, defined as
\begin{equation}\label{tfBm}
B_{\alpha,\lambda}(t)=\int_{-\infty}^{+\infty}[\textrm{e}^{-\lambda(t-x)_+}(t-x)_+^{-\alpha}
-\textrm{e}^{-\lambda(-x)_+}(-x)_+^{-\alpha}]B(\textrm{d} x),
\end{equation}
where $\lambda>0$, $\alpha<\frac{1}{2}$, the Hurst index $H=\frac{1}{2}-\alpha$, and
 \begin{equation*}
(x)_+=
\left\{
  \begin{array}{ll}
     x~~~& \hbox{for}~~x>0 \\
     0~~~& \hbox{for}~~x\leq 0.
  \end{array}
\right.
\end{equation*}
 The basic theory of tfBm is developed with application to modeling wind speed. 
Its generalized self-similarity is that for any $c>0$,
\begin{equation} \label{self-similar}
\left\{ B_{\alpha,\lambda}(ct) \right\}_{t \in \mathbb{R}} = \left\{ c^{H} B_{\alpha,c\lambda}(t) \right\}_{t \in \mathbb{R}} 
\end{equation}
in distribution and it has the covariance function
\begin{equation}\label{covariance}
 \textrm{cov}[B_{\alpha,\lambda}(t),B_{\alpha,\lambda}(s)]=\frac{1}{2}\left[C_t^2|t|^{2H}+C_s^2|s|^{2H}-C_{t-s}^2|t-s|^{2H}\right]
\end{equation}
for any $t,s\in\mathbb{R}$. The representation of $C_t^2$ is shown in (\ref{C_t^2}) with detailed derivation given in  \cite{Meerschaert:2013}. For fixed $s>0$ and large time $t$, the asymptotic behavior of this covariance is
\begin{equation}\label{tfBm_cov}
 \textrm{cov}[B_{\alpha,\lambda}(t),B_{\alpha,\lambda}(s)]\simeq \frac{1}{2}C_s^2s^{2H}+\frac{\Gamma(H+\frac{1}{2})}{(2\lambda)^{H+\frac{1}{2}}}(\textrm{e}^{\lambda s}-1)t^{H-\frac{1}{2}}\textrm{e}^{-\lambda t},
\end{equation}
on account of $K_H( t)\simeq\sqrt{\pi}(2 t)^{-\frac{1}{2}}\textrm{e} ^{ -t}$ as $t\rightarrow \infty$.
From \cite{Meerschaert:2013}, tfBm is a Gaussian process with mean value $\langle B_{\alpha,\lambda}(t)\rangle=0$ and variance $\langle B_{\alpha,\lambda}^2(t)\rangle=C_t^2|t|^{2H}$. For fixed $\lambda$ and long time $t$, the asymptotic behavior of variance is
\begin{eqnarray}\label{asym_var}
\langle B_{\alpha,\lambda}^2(t)\rangle\simeq\frac{2\Gamma{(2H)}}{(2\lambda)^{2H}}
-\frac{2\Gamma{(H+\frac{1}{2})}}{(2\lambda)^{H+\frac{1}{2}}}t^{H-\frac{1}{2}}\textrm{e}^{-\lambda t}.
\end{eqnarray}
It shows that the MSD of tfBm tends to a constant $2\Gamma{(2H)}(2\lambda)^{-2H} $ at the rate $t^{H-\frac{1}{2}} \textrm{e}^{-\lambda t}$ and thus tfBm is a localization diffusion process. In addition, from (\ref{tfBm_cov}) and (\ref{asym_var}), the correlation function of tfBm for fixed $s$ and large $t$ is a constant depending on $s$. That is to say, the correlation of the tfBm remains unchanged with the lengthening of the time interval, owing to the localization of the tfBm for long time.

Given a tfBm (\ref{tfBm}), we adopt the definition of tfGn in \cite{Chen:2017}
\begin{equation}
\gamma(t)=\frac{B_{\alpha,\lambda}(t+h)-B_{\alpha,\lambda}(t)}{h},
\end{equation}
which is similar to the definition of fractional Gaussian noise  \cite{Mandelbrot:1968}, where $h$ is small and $h\ll t$. The asymptotic behavior of its covariance function is
\begin{equation}\label{tfGnCov}
  \langle\gamma(0)\gamma(t)\rangle\simeq - \frac{\Gamma(H+1/2)\lambda^{3/2-H}}{2^{H+1/2}}\,t^{H-1/2}\textrm{e}^{-\lambda t}
\end{equation}
for fixed $\lambda$ and long times.

In the rest of this section, we introduce the time-changed tfBm by inverse $\beta$-stable subordinator, denoting as $Z(t):=B_{\alpha,\lambda}(s(t))$. 
Using the generalized self-similarity of tfBm, there exists
\begin{eqnarray}
\fl\langle Z^2(t)\rangle &=\langle B_{\alpha,\lambda}^2(s(t))\rangle =\langle s^{2H}(t) B_{\alpha,s(t)\lambda}^2(1)\rangle\nonumber\\
\fl&=\int_{-\infty}^{+\infty} \langle s^{2H}(t)[\textrm{e}^{-\lambda s(t)(1-x)_+}(1-x)_+^{-\alpha}- \textrm{e}^{-\lambda s(t)(-x)_+}(-x)_+^{-\alpha}]^2\rangle\textrm{d}x.
\end{eqnarray}
Note that unlike the subordinated fBm, here $\langle s^{2H}(t) B_{\alpha,s(t)\lambda}^2(1)\rangle$ cannot be written as $\langle s^{2H}(t)\rangle\langle B_{\alpha,s(t)\lambda}^2(1)\rangle$ on account of the dependence of $B_{\alpha,s(t)\lambda}(1)$ on $s(t)$. Therefore, for $t_1>0$ and $t_2>0$, the covariance function of $Z(t)$ is as follows
 \begin{eqnarray*}
\fl\langle Z&(t_1)Z(t_2)\rangle\\
\fl&=\langle B_{\alpha,\lambda}(s(t_1))B_{\alpha,\lambda}(s(t_2))\rangle  \\
\fl&=\frac{1}{2}\Bigg[\langle B_{\alpha,\lambda}^2(s(t_1))\rangle+\langle B_{\alpha,\lambda}^2(s(t_2))\rangle-\langle (B_{\alpha,\lambda}(s(t_1))-B_{\alpha,\lambda}(s(t_2)))^2\rangle\Bigg]  \\
\fl&=\frac{1}{2}\Bigg[\int_{-\infty}^1\langle s^{2H}(t_1)\textrm{e}^{-2\lambda(1-x)s(t_1)}\rangle (1-x)^{-2\alpha}\textrm{d}x
+\int_{-\infty}^0\langle s^{2H}(t_1)\textrm{e}^{-2\lambda(-x)s(t_1)}\rangle (-x)^{-2\alpha}\textrm{d}x\\
\fl&~~~-2\int_{-\infty}^0\langle s^{2H}(t_1)\textrm{e}^{-\lambda(1-2x)s(t_1)}\rangle (1-x)^{-\alpha}(-x)^{-\alpha}\textrm{d}x\\
\fl&~~~+\int_{-\infty}^1\langle s^{2H}(t_2)\textrm{e}^{-2\lambda(1-x)s(t_2)}\rangle (1-x)^{-2\alpha}\textrm{d}x
+\int_{-\infty}^0\langle s^{2H}(t_2)\textrm{e}^{-2\lambda(-x)s(t_2)}\rangle (-x)^{-2\alpha}\textrm{d}x\\
\fl&~~~-2\int_{-\infty}^0\langle s^{2H}(t_2)\textrm{e}^{-\lambda(1-2x)s(t_2)}\rangle (1-x)^{-\alpha}(-x)^{-\alpha}\textrm{d}x\\
\fl&~~~-\int_{-\infty}^1\langle |s(t_1)-s(t_2)|^{2H}\textrm{e}^{-2\lambda(1-x)|s(t_1)-s(t_2)|}\rangle (1-x)^{-2\alpha}\textrm{d}x\\
\fl&~~~-\int_{-\infty}^0\langle |s(t_1)-s(t_2)|^{2H}\textrm{e}^{-2\lambda(-x)|s(t_1)-s(t_2)|}\rangle (-x)^{-2\alpha}\textrm{d}x \\
\fl&~~~+2\int_{-\infty}^0\langle |s(t_1)-s(t_2)|^{2H}\textrm{e}^{-\lambda(1-2x)|s(t_1)-s(t_2)|}\rangle (1-x)^{-\alpha}(-x)^{-\alpha}\textrm{d}x\Bigg].
\end{eqnarray*}
Combining $\mathcal{L}_{t\rightarrow u}[f(s,t)]=u^{\beta-1}\textrm{e}^{-su^\beta}$ with (\ref{twotimepdf}), the covariance function of subordinated tfBm $Z(t)$ in Laplace space ($t_1\rightarrow u_1$, $t_2\rightarrow u_2$) is
\begin{eqnarray*}
\langle Z(u_1)Z(u_2)\rangle
=&\frac{u_2^\beta}{u_1(u_1+u_2)^\beta}\int_{-\infty}^0g_1(u_2; -2x)(-x)^{-2\alpha}\textrm{d}x\\
&+\frac{u_1^\beta}{u_2(u_1+u_2)^\beta}\int_{-\infty}^0g_1(u_1; -2x)(-x)^{-2\alpha}\textrm{d}x\\
&
-\frac{u_2^\beta}{u_1(u_1+u_2)^\beta}\int_{-\infty}^0g_1(u_2; 1-2x)(1-x)^{-\alpha}(-x)^{-\alpha}\textrm{d}x\\
&-\frac{u_1^\beta}{u_2(u_1+u_2)^\beta}\int_{-\infty}^0g_1(u_1; 1-2x)(1-x)^{-\alpha}(-x)^{-\alpha}\textrm{d}x,
\end{eqnarray*}
where $g_1(u;x)=\frac{\Gamma(2H+1)u^{\beta-1}}{(\lambda x+u^\beta)^{2H+1}}$ and $\alpha<\frac{1}{2}$. By the inverse Laplace transform, we finally obtain the covariance function of the time-changed tfBm $Z(t)$:
\begin{eqnarray*}
\fl\langle Z&(t_1)Z(t_2)\rangle \\
\fl&=\Theta(t_1-t_2)\frac{\Gamma(2H)}{(2\lambda)^{2H}}\left[1+\frac{1}{\Gamma(1-\beta)\Gamma(\beta)}B\left(\beta, 1-\beta; \frac{t_2}{t_1}\right)\right]
-\Theta(t_1-t_2)\int_0^{t_2}g_2(t')\textrm{d}t'\\[3pt]
\fl&-\Theta(t_1-t_2)\left[\int_0^{t_2}g_2(t_1-t')\textrm{d}t'
+\frac{1}{\Gamma(\beta)\Gamma(1-\beta)}\int_{t_2}^{t_1}B\left(\beta, 1-\beta; \frac{t_2}{t'}\right)g_2(t_1-t')\textrm{d}t'\right]\\[3pt]
\fl&+\Theta(t_2-t_1)\frac{\Gamma(2H)}{(2\lambda)^{2H}}\left[1+\frac{1}{\Gamma(1-\beta)\Gamma(\beta)}B\left(\beta, 1-\beta; \frac{t_2}{t_1}\right)\right]
-\Theta(t_2-t_1)\int_0^{t_1}g_2(t')\textrm{d}t'\\[3pt]
\fl&-\Theta(t_2-t_1)\left[\int_0^{t_1}g_2(t_2-t')\textrm{d}t'
+\frac{1}{\Gamma(\beta)\Gamma(1-\beta)}\int_{t_1}^{t_2}B\left(\beta, 1-\beta; \frac{t_2}{t'}\right)g_2(t_2-t')\textrm{d}t'\right],
\end{eqnarray*}
with
\begin{eqnarray*}
g_2(t)=\Gamma(2H+1)t^{2H\beta-1}\int_{-\infty}^0 E_{\beta,2H\beta}^{2H+1}(-\lambda(1-2x)t^\beta)(1-x)^{-\alpha}(-x)^{-\alpha}\textrm{d}x,
\end{eqnarray*}
and its Laplace transform is
\begin{eqnarray*}
\mathcal{L}_{t\rightarrow u}[g_2(t)]= \int_{-\infty}^0 \frac{\Gamma(2H+1) u^\beta}{[\lambda (1-2x)+u^\beta]^{2H+1}}(1-x)^{-\alpha}(-x)^{-\alpha}\textrm{d}x,
\end{eqnarray*}
where $E_{\alpha,\beta}^{\delta}(z)=\Sigma_{k=0}^\infty\frac{\Gamma(\delta+k)}{\Gamma(\delta)\Gamma(\alpha k+\beta)}\frac{z^k}{k!}$ is the three parameter Mittag-Leffler function \cite{Prabhakar:1971,Liemert:2017,Sandev:2017}.
When $t_1=t_2$, one can get the expression of the variance:
\begin{eqnarray*}
\langle Z^2(t)\rangle=\frac{2\Gamma(2H)}{(2\lambda)^{2H}}-2\int_0^tg_2(t')\textrm{d}t'.
\end{eqnarray*}
For long times, the asymptotic behavior of the variance $\langle Z^2(t)\rangle$ is
\begin{equation}\label{Z^2}
\langle Z^2(t)\rangle\simeq2\Gamma(2H)(2\lambda)^{-2H}-Gt^{-\beta},
\end{equation}
with $G=\frac{2\Gamma(2H+1)}{\Gamma(1-\beta)}\int_{-\infty}^0\frac{1}{[\lambda(1-2x)]^{2H+1}}
(1-x)^{-\alpha}(-x)^{-\alpha}\textrm{d}x$ and $0<\beta<1$, which means that the subordinated tfBm is also a localization diffusion process.

\begin{figure}[!htb]
\flushright
\begin{minipage}{0.8\linewidth}
  \centerline{\includegraphics[width=10cm,height=6.5cm]{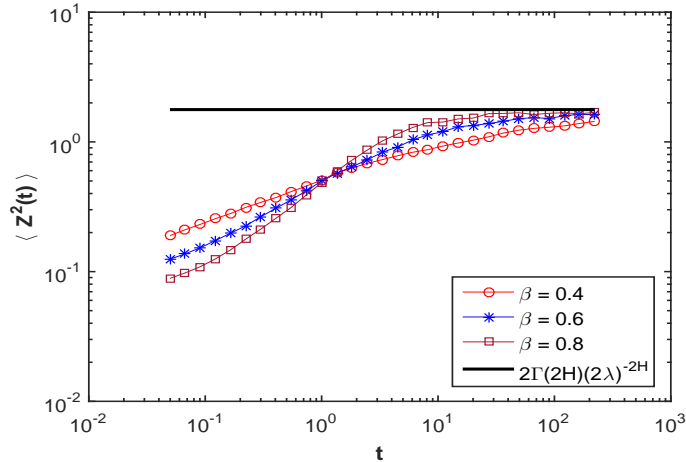}}
  \end{minipage}
\caption{Simulation results of the variance of subordinated tfBm. Parameter values: $H=0.7$, $\lambda=0.5$, and the number of simulation trajectories is $3000$.}\label{sub_tfBm_variance}
\end{figure}

\begin{figure}[!htb]
\flushright
\begin{minipage}{0.8\linewidth}
  \centerline{\includegraphics[width=10cm,height=6.5cm]{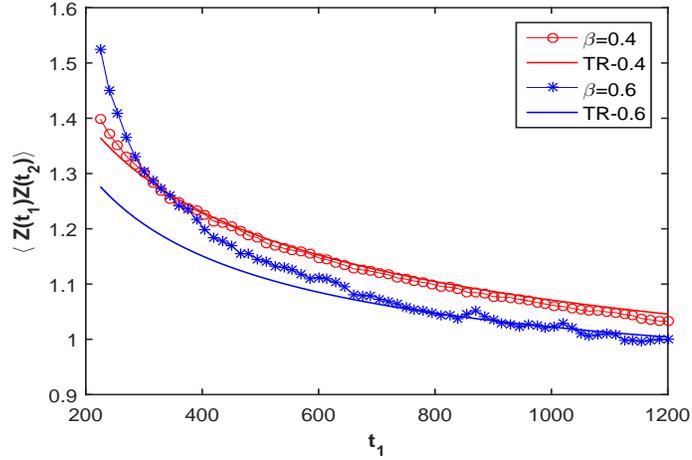}}
  \end{minipage}
\caption{Covariance function (\ref{cov_sub}) of subordinated tfBm and the corresponding simulation results represented by marks. Solid red line ($TR-0.4$) represents the theoretical result with $\beta=0.4$ and the solid blue line ($TR-0.6$) is the theoretical result with $\beta=0.6$. Parameter values: $H=0.7$, $\lambda=0.5$, $t_2=200$, and the number of trajectories is $16000$.}\label{sub_tfBm_covariance}
\end{figure}

The simulation results of the variance (\ref{Z^2}) are shown in figure \ref{sub_tfBm_variance} and it can be noted that the variance (\ref{Z^2}) tends to the constant $2\Gamma(2H)(2\lambda)^{-2H}$ with the speed of $t^{-\beta}$. Especially for the case $\beta=1$, the variance of the subordinated tfBm  $Z(t)$ is $\langle Z^2(t)\rangle=C_t^2t^{2H}$ ,
which is obtained by using that when $\beta=1$, there is
\begin{eqnarray*}
\fl\int_0^t g_2(t')\textrm{d}t' =\int_{-\infty}^0 \textrm{e}^{-\lambda(1-2x)t}t^{2H}(1-x)^{-\alpha}(-x)^{-\alpha}\textrm{d}x=
\frac{\Gamma{(H+1/2)}t^H}{\sqrt{\pi}(2\lambda)^H}K_H(\lambda t).
\end{eqnarray*}
So that for the case $\beta=1$,  $\langle Z^2(t)\rangle$ is consistent with the variance of tfBm $\langle B_{\alpha,\lambda}^2(t)\rangle$ .
For fixed $t_2$ and large $t_1$, the asymptotic behavior of the covariance function $\langle Z(t_1)Z(t_2)\rangle$ behaves as
\begin{eqnarray}\label{cov_sub}
\langle Z(t_1)Z(t_2)\rangle&\simeq \frac{\Gamma(2H)}{(2\lambda)^{2H}}
-\int_0^{t_2}g_2(t')\textrm{d}t'\nonumber
+\frac{\Gamma(2H)}{\Gamma(1-\beta)\Gamma(1+\beta)(2\lambda)^{2H}} \,t_2^\beta t_1^{-\beta}\nonumber\\
&=\frac{1}{2}\langle Z^2(t_2)\rangle+\frac{\Gamma(2H)}{\Gamma(1-\beta)\Gamma(1+\beta)(2\lambda)^{2H}} \,t_2^\beta t_1^{-\beta}.
\end{eqnarray}
The corresponding simulation results are shown in figure \ref{sub_tfBm_covariance}; these curves are consistent with the theoretical results (\ref{cov_sub}) for large time $t_1$. One can see that the covariance function tends to $\frac{1}{2}\langle Z^2(t_2)\rangle$ at the rate $t_1^{-\beta}$, which shows the long-range dependence of the subordinated tfBm.

Comparing the subordinated tfBm $Z(t)$ with the original tfBm $B_{\alpha, \lambda}(t)$, it can be noted that both the variances of the two processes tend to a constant $2\Gamma(2H)(2\lambda)^{-2H}$ for long time limit, but the speed is different.
The former is with $t^{-\beta}$, independent of the Hurst index $H$, while the latter is with $t^{H-1/2}\textrm{e}^{-\lambda t}$, which implies that the method of subordination slows down the speed of converging to the final state.
Besides that, for fixed $t_2$ and large $t_1$, the covariances of $Z(t)$ and $B_{\alpha, \lambda}(t)$ all tend to half of their variance, i.e., $\frac{1}{2}\langle Z^2(t_2)\rangle$ and $\frac{1}{2}\langle B_{\alpha, \lambda}^2(t_2)\rangle$, except the difference in speed (one is with $t^{-\beta}$, and another one is with $t^{H-\frac{1}{2}}\textrm{e}^{-\lambda t}$). We know that tfBm is an ergodic process \cite{Chen:2017}, while the time-changed tfBm by inverse $\beta$-stable subordinator is non-ergodic. 

Next, we consider the covariance function of the increments of subordinated tfBm $Z(t)$, denoted as $Y_t=Z(t+h)-Z(t)$ for fixed small $h$. By the above method and the asymptotic behavior (\ref{cov_sub}) of $\langle Z(t_1)Z(t_2)\rangle$, the asymptotic expression of the covariance function of $Y_t$ for fixed $h$ and long time $t$ is
\begin{eqnarray*}
\fl\langle Y_0Y_t\rangle&=\langle Z(h)Z(t+h)\rangle-\langle Z(h)Z(t)\rangle
\simeq -\frac{\Gamma(2H)}{(2\lambda)^{2H}\Gamma(1-\beta)\Gamma(\beta)}h^{\beta+1}t^{-\beta-1}.
\end{eqnarray*}
This means that the covariance function $\langle Y_0Y_t\rangle$ of the increment of subordinated process $Z(t)$ tends to zero at the rate $t^{-\beta-1}$, while the covariance function $\langle\gamma(0)\gamma(t)\rangle$ of the increment of tfBm $B_{\alpha, \lambda}(t)$ approaches to zero at the rate $t^{H-\frac{1}{2}}\textrm{e}^{-\lambda t}$ in (\ref{tfGnCov}).
Performing the $\beta$-stable subordination, the obtained new process $Z(t)$ is long-range dependent but with  short-range dependent increments, and hence this process may possibly model some financial data \cite{Scalas:2006} in real  applications.

\section{Conclusion}\label{four}
The tfBm was recently introduced, which can effectively describe wind speed. This paper further considers the time-changed non-Markovian Langevin systems, including  time-changed fLe, time-changed tfLe, and time-changed tfBm, with potential applications in finance, biology, and physics.
Through the standard approach of subordination, we explicitly discuss the diffusion types, moments, Klein-Kramers equation, and correlation structures of the subordinated tfLe with inverse $\beta$-stable process. An interesting phenomenon is observed, i.e., 
the subordinated tfLe can undergo subdiffusion or superdiffusion, even normal diffusion, depending on the value of $\beta$.
The MSD of the subordinated tfLe is analogous to the case of the time-changed Langevin equation with biasing external force, implying a similar superdiffusion. But the mechanisms are completely different. The former mainly results from the power-law distributed waiting time of which the occasional immobilization slows down the original process (ballistic diffusion),  while the latter stems from the convection term, where the external biasing force acts only at the time of the jumps but not affects the dynamics of the diffusing particle during the waiting periods, slowing down the center mass of the particles. 
 For the time-changed tfBm by inverse $\beta$-stable subordinator, though the variance and covariance are still a constant for the long time case, the speed of approaching the final state is slower than the original process. More specifically, the converging speed of the original process is $t^{H-\frac{1}{2}}\textrm{e}^{-\lambda t}$, while the one of the subordinated process is $t^{-\beta}$, being independent on the Hurst index $H$.

\section*{Acknowledgments}
This work was supported by the National Natural Science Foundation of China under grant no. 11671182, and the Fundamental Research Funds for the Central Universities under grants no. lzujbky-2018-ot03 and no. lzujbky-2017-ot10.


\end{document}